# Antitrust and Artificial Intelligence (AAI):
# The Antitrust Vigilance Lifecycle And AI Legal Reasoning Autonomy

**Dr. Lance B. Eliot**
Chief AI Scientist, Techbruim; Fellow, CodeX: Stanford Center for Legal Informatics
Stanford, California, USA

## Abstract

There is an increasing interest in the entwining of the field of antitrust with the field of Artificial Intelligence (AI), frequently referred to jointly as Antitrust and AI (AAI) in the research literature. This study focuses on the synergies entangling antitrust and AI, doing so to extend the literature by proffering the primary ways that these two fields intersect, consisting of: (1) the application of antitrust to AI, and (2) the application of AI to antitrust. To date, most of the existing research on this intermixing has concentrated on the former, namely the application of antitrust to AI, entailing how the marketplace will be altered by the advent of AI and the potential for adverse antitrust behaviors arising accordingly. Opting to explore more deeply the other side of this coin, this research closely examines the application of AI to antitrust and establishes an *antitrust vigilance lifecycle* to which AI is predicted to be substantively infused for purposes of enabling and bolstering antitrust detection, enforcement, and post-enforcement monitoring. Furthermore, a gradual and incremental injection of AI into antitrust vigilance is anticipated to occur as significant advances emerge amidst the Levels of Autonomy (LoA) for AI Legal Reasoning (AILR).

## 1 Background on Antitrust and AI

In Section 1 of this paper, the literature on Antitrust and AI (AAI) is introduced and addressed. Doing so establishes groundwork for the subsequent sections. Section 2 introduces the Levels of Autonomy (LoA) of AI Legal Reasoning (AILR), which is instrumental in the discussions undertaken in Section 3. This provides an expanded viewpoint of AAI, including articulating the primary way that these two fields intertwine, along with proffering an in-depth exploration of the application of AI to antitrust, along with other vital facets. Section 4 provides various additional research implications and anticipated impacts upon salient practice-of-law considerations.

This paper then consists of these four sections:

- Section 1: Background on Antitrust and AI
- Section 2: Levels of Autonomy (LOA) of AI Legal Reasoning (AILR)
- Section 3: Focus on AI Applied to Antitrust
- Section 4: Additional Considerations and Future Research

### 1.1 Overview of Antitrust and AI

An expanding field of substantive interest for the theory of the law and the practice-of-law entails Antitrust and AI (commonly referred to as AAI). To some degree, there are expressed qualms that the antitrust community has been slow to adopt and consider the ramifications of AI [18] [34] [47] [69] [77] [81].

In Section 3, we address the aspect that there are two major forms of synergy between antitrust and AI, consisting of the application of antitrust to AI, and the application of AI to antitrust.

Per the research by Deng [18], attention to-date has appeared to primarily concentrate on the application of antitrust to AI. This includes the overarching concern that businesses adopting AI will be able to collude in a manner and degree not previously envisioned, and for which then promotes and accelerates antitrust behaviors in the marketplace.

According to Deng [18], there are at least two avenues for this kind of considered anticompetitive behavior as prodded via the use of AI:

> "In the antitrust community, the recent interest in AI is also driven in part by concerns about algorithmic collusion. At least two ways in which computer algorithms could facilitate collusion have been identified. First, computer algorithms could be used to implement a price-fixing agreement, e.g., an agreed-upon price or production level or automating the detection of "cheating" and retaliation."

And the other means is [18]:

> "This is an important observation because the antitrust community is also concerned with another type of much more sophisticated algorithmic collusion, i.e., the possibility that algorithms could ultimately learn to collude without human facilitation."

A crucial meta-perspective on this preceding point is that the AI can either be established in a means to be directed by human hands while undertaking antitrust behaviors, plus can be ostensibly *let loose* via the inclusion of Machine Learning and Deep Learning to perform antitrust behaviors without a human hand at the wheel per se.

In the research by Petit [62], there is an indication that this kind of antitrust behavior can be characterized in a threefold manner:

> "First, algorithms will widen instances in which known forms of anticompetitive conduct occurs. The AAI scholarship conjectures that express and tacit collusion, as well as almost perfect behavioral discrimination, will be more common.
>
> Second, algorithmic markets will display new forms of anticompetitive conduct in non-price dimensions like data capture, extraction, and co-opetition (between 'super-platforms' and applications developers) which challenge established antitrust doctrine.
>
> Third, deception is a design feature of algorithmic markets. Behind the 'façade' of competition, consumers are nudged in exploitative transactions."

When considering how a government will respond to these likely AI-enabled antitrust behaviors, the work of Hayashi and Arai [44] asserts that governmental agencies and laws will need to be reconsidered and possibly revamped, doing so to try and stabilize the marketplace playing field into one of fair and balanced competition in light of the leveraging of AI.

One offshoot of the AAI realm involves the thorny question of legal liability and legally understood notions of intent when AI that is based on Machine Learning and Deep Learning is being used in seemingly antitrust-related ways [36]:

> "The consideration of the 'autonomous agent' raises ethical and policy questions on the relationship between humans and machines. In such instances, can the law attribute liability to companies for their computers' actions? At what stage, if any, would the designer or operator relinquish responsibility over the acts of the machine? Evidently, defining a benchmark for illegality in such cases is challenging. It requires close consideration of the relevant algorithm to establish whether any illegal action could have been anticipated or was predetermined. Such review requires consideration of the programming of the machine, available safeguards, its reward structure, and the scope of its activities. The ability to identify the strand which is of direct relevance is questionable. The complexity of the algorithms' data-processing and self-learning increases the risk that enforcers, in undertaking this daunting undertaking, stray far afield of rule of ideals, such as transparency, objectivity, predictability, and accuracy. Further, one must consider the extent to which humans may truly control self-learning machines."

Another variant of the focus on antitrust as applied to AI consists of the dimension that there is AI innovation that occurs in a marketplace, and a thriving AI developer ecosystem is presumed to be vital to the ongoing advancement of AI-based technologies and capabilities.

This is another point of concern in the AAI community since there is a possibility that antitrust enforcement could be applied toward the mechanisms of the AI innovation ecosystem. This use of antitrust could be negatively disruptive to AI advancements and seemingly dampen or perhaps curtail pending new AI-based capabilities.

Foster and Zachary [38] emphasize that besides the qualms of how this might adversely impact industry and the ability of businesses to employ and deploy AI, it could have a severe indirect and consequential impact on military defense readiness too:

> "Unlike with many prior defense technologies, the private sector currently drives the development of AI. Therefore, to use AI to America's national security advantage, the Pentagon will rely in large part on the domestic private-sector AI ecosystem. At the same time, antitrust policymakers are contemplating significant changes to this ecosystem, and are even considering breaking up its largest companies. How would such an action affect the Pentagon's AI capabilities?"

By-and-large, much of the research on AAI oftentimes takes a particular perspective of either the considered downside or adverse element of antitrust as applied to AI, or on other occasions takes the opposite position that antitrust could potentially further enable AI advancement and adoption in the marketplace (though, this latter perspective is a somewhat rarer argued posture).

In research by Rab [64], he cautions that we ought to be giving attention to both sides of the coin, as it were, and not allow ourselves to fall into the trap that one side or the other does not exist or is somehow axiomatically unworthy of balanced attention:

> "The AI antitrust scholarship makes a bold claim that AI is an enabler of tacit collusion and could increase the scope for anti-competitive outcomes at even lower levels of concentration than associated with antitrust orthodoxy. However, even the brief examination of these claims in this article has unearthed alternative hypotheses which need to be fully tested before the theory can be incorporated in policy and legal environments without running the risk of being counter-productive."

### 1.2 Complexities and Vagaries of Antitrust

Though antitrust might seemingly be a straightforward notion, the reality is that antitrust bears a great deal of complexity, vagary, and altogether defies any inarguable definitive scope and ironclad meaning.

The simplest means to construe antitrust can be readily expressed in this manner as stated by the U.S. Department of Justice (DOJ) [82]:

> "The U.S. antitrust laws represent the legal embodiment of our nation's commitment to a free market economy in which the competitive process of the market ensures the most efficient allocation of our scarce resources and the maximization of consumer welfare."

Furthermore, the antitrust process to presumably carry out that mission is indicated in this way by the DOJ:

> "The Antitrust Division's mission is to promote economic competition through enforcing and providing guidance on antitrust laws and principles. When it comes to enforcement, the Division's policy, in general, is to proceed by criminal investigation and prosecution in cases involving horizontal, "per se" unlawful agreements such as price fixing, bid rigging, and market allocation. Civil process and, if necessary, civil prosecution is used with respect to other suspected antitrust violations, including those that require analysis under the "rule of reason," as well as some offenses that historically have been labeled "per se" by the courts. There are a number of situations where, although the conduct may appear to be a "per se" violation of law, criminal investigation or prosecution may not be appropriate. These situations may include cases in which (1) the case law is unsettled or uncertain; or (2) there are truly novel issues of law or fact presented."

There is a kind of triad associated with the antitrust efforts in the U.S. [49], consisting of (1) the laws that are devised by the courts, (2) the prosecutorial efforts of the DOJ and similarly antitrust-tasked agencies, and (3) the funding to those government agencies for the purposes of taking on antitrust activities including antitrust enforcement.

Consider how the triad interacts amongst the three components, akin to a three-legged stool that requires all three legs, demonstrably aligned to be able to stand upright.

If there are insufficient laws or laws that are seen as weak toward antitrust behaviors, presumably the triad is accordingly undercut. If the government agencies tasked with antitrust opt to not pursue antitrust behaviors, even if there are sufficient laws to warrant such pursuit, this presumably undercuts the antitrust policing. And, even if the government agencies are desirous of such pursuits, and the laws are in place, without adequate antitrust-related funding the result is a mere hollow aspiration waiting to be fulfilled.

As Kades at Yale University stated [49]:

> "The effectiveness of the U.S. antitrust laws in protecting competition depends on the three key factors. The first is jurisprudential doctrines that courts develop. The second is the prosecutorial discretion that enforcers—the Antitrust Division of the Department of Justice, the Federal Trade Commission, and state attorneys general—employ. And the third is the fiscal resources provided to the enforcers. It can be difficult to disentangle the role of these factors. The federal government, for example, may bring fewer antitrust cases because it has changed its enforcement philosophy. Or a judicial decision may limit the reach of the antitrust laws by limiting the government's ability to challenge certain types of cases. Similarly, a change in enforcement discretion or the courts broadening the scope of the antitrust law could lead to increased enforcement. Indirectly, judicial or evidentiary rules that increase the cost of successfully pursuing cases can reduce the number of antitrust cases (and the reverse could increase it). Increasing or decreasing appropriations for the antitrust enforcement agencies also can affect both the degree of antitrust enforcement and its impact."

A core deficiency underlying the entire premise of antitrust is that there is an axiomatic means to ascertain whether a marketplace is being anti-competitively overrun. Not everyone concurs that there is some form of calculus or infallible method of ascertaining the competitive versus anti-competitive conjecture.

Per the rather sharp toothed remarks of Stigler [73]:

> "Economists have their glories, but I do not believe that antitrust law is one of them."

This claim, which some might find alarming, or possibly unfounded, he supports via this indication [73]:

> "In a series of studies done in the early 1970s, economists assumed that important losses to consumers from limits on competition existed, and constructed models to identify the markets where these losses would be greatest. Then they compared the markets where government was enforcing antitrust laws with the markets where governments *should* enforce the laws if consumer well-being was the government's paramount concern. The studies concluded unanimously that the size of consumer losses from monopoly played little or no role in government enforcement of the law."

There is a lot at stake when considering that the very act or activity of potential governmental intervention is seemingly bereft of preciseness and not amenable to widespread understanding, thus putting businesses into the unenviable position of presumably not being able to gauge what the rules of the game really are (with respect to what is deemed as antitrust behavior).

Consider these salient points about how the ambiguity of what antitrust consists of can be both a blessing and a curse to the marketplace [20]:

> "A fundamental difficulty facing the court is the incommensurability of the stakes. If the court errs by condemning a beneficial practice, the benefits may be lost for good. Any other firm that uses the condemned practice faces sanctions in the name of stare decisis, no matter the benefits. If the court errs by permitting a deleterious practice, though, the welfare loss decreases over time. The legal system should be designed to minimize the total costs of (1) anticompetitive practices that escape condemnation; (2) competitive practices that are condemned or deterred; and (3) the system itself. The third is easiest to understand. Some practices, although anticompetitive, are not worth deterring. We do not hold three-week trials about parking tickets. And when we do seek to deter, we want to do so at the least cost."

All told, per [20] the assumption that antitrust is a handy or helpful tool is raft with falsehood since the antitrust looming threat of enforcement can undermine the marketplace, just as anti-competitive behavior can undermine the marketplace:

"Antitrust is an imperfect tool for the regulation of competition. Imperfect because we rarely know the right amount of competition there should be, because neither judges nor juries are particularly good at handling complex economic arguments, and because many plaintiffs are interested in restraining rather than promoting competition."

### 1.3 Due Process and Antitrust

If antitrust is potentially an ill-devised tool or at least an imperfect one, the question of due process and fairness certainly enters into the picture.

First, consider the importance of the rule of law [88]:

"The rule of law is one of the founding principles of the United States of America and has shaped U.S. legal thinking and practice in many areas. The basic idea is simple: the people and their actions are not governed and regulated by arbitrary decisionmakers, but by a set of rules that serves as a check against potential abuses of power."

And, the rule of law is especially vital when considering antitrust enforcement [88]:

"Due process and fairness are particularly important in antitrust enforcement. Because the U.S. was the first country to enact an antitrust law, it has enjoyed the greatest opportunity to develop its enforcement practices."

One perspective is that the government will act in a rationalized and non-politically motivated way, presumably rooting out antitrust behaviors in an unbiased manner. This belies though the reality that the politics of the day can readily be interjected into antitrust oversight and enforcement actions.

Indeed, the recent work by Schrepel points out that to some extent there is an emerging semblance of romanticizing of antitrust, pitting the technology elites against the public-at-large, doing so in a politically divined shroud [71]:

"Increasing romanticization could critically jeopardize decades of jurisprudential construction, causing economic disruption, destabilization of the law, and blindness towards real anti-competitive practices on the part of antitrust authorities, consequently placing the rule of law at risk."

It is altogether conceivable and perhaps patently obvious to some that the antitrust notion can be wielded as a somehow neutralized axiom that seeks to optimize competition and prevent the usurping of a marketplace by antitrust behaviors, meanwhile, the antitrust arm-twisting power-punching capacity can be exercised by politicians that have a particular political gain in mind.

Succinctly stated by McChesney [57]:

"If public-interest rationales do not explain antitrust, what does? A final set of studies has shown empirically that patterns of antitrust enforcement are motivated at least in part by political pressures unrelated to aggregate economic welfare. For example, antitrust is useful to politicians in stopping mergers that would result in plant closings or job transfers in their home districts."

Yet another consideration of the efficacy of antitrust as a governmental tool involves the nature of innovation itself and how it fares or evolves in the marketplace. One argument to be made is that AI, in particular, needs as much latitude as possible at this time to properly percolate and mature, for which a heavy-handed antitrust clash might disturb (this was earlier pointed out in the remarks about the Pentagon and reliance on AI innovation arising from industry).

A cornerstone research paper by Schrepel [72], defines and proffers that we need to encapsulate predatory innovation into the vernacular of antitrust law:

"In fact, the terms of *predatory innovation*—which the author defines as the alteration of one or more technical elements of a product to limit or eliminate competition—describes all practices that, under the

> guise of real innovations, are anti-competitive strategies aimed at eliminating competition without benefiting consumers."

This provides yet another important consideration when weighing the antitrust triad and how it proceeds.

### 1.4 Applying AI to Antitrust

The preceding subsections have focused on the realm of AAI that deals with the application of antitrust to AI, and as explained earlier in Subsection 1.1, ostensibly has received the preponderance of attention in the AAI research literature to-date.

Shifting the focus to the other side of the coin, we next consider the application of AI to antitrust. This will also become the bulk of the attention in Section 3 and as based on the contextual background provided in Section 2.

A recent study by Engstrom, Ho, Sharkey, and Cuellar examined how AI is being used by federal agencies [35]:

> "Artificial intelligence (AI) promises to transform how government agencies do their work. Rapid developments in AI have the potential to reduce the cost of core governance functions, improve the quality of decisions, and unleash the power of administrative data, thereby making government performance more efficient and effective. Agencies that use AI to realize these gains will also confront important questions about the proper design of algorithms and user interfaces, the respective scope of human and machine decision-making, the boundaries between public actions and private contracting, their own capacity to learn over time using AI, and whether the use of AI is even permitted."

There have been sparse and spotty efforts to incorporate AI into governmental antitrust efforts. This is perhaps handy since it implies a rife potential for an evergreen approach to applying AI into antirust. At the same time, the adoption of AI into antitrust efforts could be undertaken with haphazard attention, producing either mismanaged results or worse still leading to antitrust that is overtaken by unsavory AI.

There is an ongoing concern that AI is at times implemented without proper controls and monitoring. Qualms too are that AI in governmental agencies might be lacking in explainability or interpretability (having AI that can explain its actions is typically referred to as XAI, see [21] [22] [23]).

A well-taken advisory caution by [35] emphasizes the importance of properly applying AI to governmental uses:

> "Managed well, algorithmic governance tools can modernize public administration, promoting more efficient, accurate, and equitable forms of state action. Managed poorly, government deployment of AI tools can hollow out the human expertise inside agencies with few compensating gains, widen the public-private technology gap, increase undesirable opacity in public decision-making, and heighten concerns about arbitrary government action and power. Given these stakes, agency administrators, judges, technologists, legislators, and academics should think carefully about how to spur government innovation involving the appropriate use of AI tools while ensuring accountability in their acquisition and use."

Section 2 will introduce the principles underlying AI Legal Reasoning and the various levels of autonomy therein. Section 3 then continues this herein discussion about applying AI to antitrust and does so in light of the aforementioned cautionary insights about the adoption of AI in governmental agencies.

## 2 Levels of Autonomy (LOA) of AI Legal Reasoning (AILR)

In this section, a framework for the autonomous levels of AI Legal Reasoning is summarized and is based on the research described in detail in Eliot [23] [24] [25] [26] [27] [28] [29].

These autonomous levels will be portrayed in a grid that aligns with key elements of autonomy and as matched to AI Legal Reasoning. Providing this context will be useful to the later sections of this paper and will be utilized accordingly.

The autonomous levels of AI Legal Reasoning are as follows:

Level 0: No Automation for AI Legal Reasoning

Level 1: Simple Assistance Automation for AI Legal Reasoning

Level 2: Advanced Assistance Automation for AI Legal Reasoning

Level 3: Semi-Autonomous Automation for AI Legal Reasoning

Level 4: Domain Autonomous for AI Legal Reasoning

Level 5: Fully Autonomous for AI Legal Reasoning

Level 6: Superhuman Autonomous for AI Legal Reasoning

## 2.1 Details of the LoA AILR

See **Figure A-1** for an overview chart showcasing the autonomous levels of AI Legal Reasoning as via columns denoting each of the respective levels.

See **Figure A-2** for an overview chart similar to Figure A-1 which alternatively is indicative of the autonomous levels of AI Legal Reasoning via the rows as depicting the respective levels (this is simply a reformatting of Figure A-1, doing so to aid in illuminating this variant perspective, but does not introduce any new facets or alterations from the contents as already shown in Figure A-1).

### 2.1.1 Level 0: No Automation for AI Legal Reasoning

Level 0 is considered the no automation level. Legal reasoning is carried out via manual methods and principally occurs via paper-based methods.

This level is allowed some leeway in that the use of say a simple handheld calculator or perhaps the use of a fax machine could be allowed or included within this Level 0, though strictly speaking it could be said that any form whatsoever of automation is to be excluded from this level.

### 2.1.2 Level 1: Simple Assistance Automation for AI Legal Reasoning

Level 1 consists of simple assistance automation for AI legal reasoning.

Examples of this category encompassing simple automation would include the use of everyday computer-based word processing, the use of everyday computer-based spreadsheets, access to online legal documents that are stored and retrieved electronically, and so on.

By-and-large, today's use of computers for legal activities is predominantly within Level 1. It is assumed and expected that over time, the pervasiveness of automation will continue to deepen and widen, and eventually lead to legal activities being supported and within Level 2, rather than Level 1.

### 2.1.3 Level 2: Advanced Assistance Automation for AI Legal Reasoning

Level 2 consists of advanced assistance automation for AI legal reasoning.

Examples of this notion encompassing advanced automation would include the use of query-style Natural Language Processing (NLP), Machine Learning (ML) for case predictions, and so on.

Gradually, over time, it is expected that computer-based systems for legal activities will increasingly make use of advanced automation. Law industry technology that was once at a Level 1 will likely be refined, upgraded, or expanded to include advanced capabilities, and thus be reclassified into Level 2.

### 2.1.4 Level 3: Semi-Autonomous Automation for AI Legal Reasoning

Level 3 consists of semi-autonomous automation for AI legal reasoning.

Examples of this notion encompassing semi-autonomous automation would include the use of Knowledge-Based Systems (KBS) for legal reasoning, the use of Machine Learning and Deep Learning (ML/DL) for legal reasoning, and so on.

Today, such automation tends to exist in research efforts or prototypes and pilot systems, along with some commercial legal technology that has been infusing these capabilities too.

### 2.1.5 Level 4: Domain Autonomous for AI Legal Reasoning

Level 4 consists of domain autonomous computer-based systems for AI legal reasoning.

This level reuses the conceptual notion of Operational Design Domains (ODDs) as utilized in the autonomous vehicles and self-driving cars' levels of autonomy, though in this use case it is being applied to the legal domain [24] [25] [26] [27]. Essentially, this entails any AI legal reasoning capacities that can operate autonomously, entirely so, but that is only able to do so in some limited or constrained legal domain.

### 2.1.6 Level 5: Fully Autonomous for AI Legal Reasoning

Level 5 consists of fully autonomous computer-based systems for AI legal reasoning.

In a sense, Level 5 is the superset of Level 4 in terms of encompassing all possible domains as per however so defined ultimately for Level 4. The only constraint, as it were, consists of the facet that the Level 4 and Level 5 are concerning human intelligence and the capacities thereof. This is an important emphasis due to attempting to distinguish Level 5 from Level 6 (as will be discussed in the next subsection)

It is conceivable that someday there might be a fully autonomous AI legal reasoning capability, one that encompasses all of the law in all foreseeable ways, though this is quite a tall order and remains quite aspirational without a clear-cut path of how this might one day be achieved. Nonetheless, it seems to be within the extended realm of possibilities, which is worthwhile to mention in relative terms to Level 6.

### 2.1.7 Level 6: Superhuman Autonomous for AI Legal Reasoning

Level 6 consists of superhuman autonomous computer-based systems for AI legal reasoning.

In a sense, Level 6 is the entirety of Level 5 and adds something beyond that in a manner that is currently ill-defined and perhaps (some would argue) as yet unknowable. The notion is that AI might ultimately exceed human intelligence, rising to become superhuman, and if so, we do not yet have any viable indication of what that superhuman intelligence consists of and nor what kind of thinking it would somehow be able to undertake.

Whether a Level 6 is ever attainable is reliant upon whether superhuman AI is ever attainable, and thus, at this time, this stands as a placeholder for that which might never occur. In any case, having such a placeholder provides a semblance of completeness, doing so without necessarily legitimatizing that superhuman AI is going to be achieved or not. No such claim or dispute is undertaken within this framework.

## 3 Focus on AI Applied to Antitrust

In this Section 3, various aspects of Antitrust and Artificial Intelligence (AAI) will be identified and discussed, particularly with respect to two key elements: (1) Establishing an AAI Antitrust Vigilance Lifecycle, and (2) AAI and the Levels of Autonomy in AI Legal Reasoning (AILR).

A series of diagrams and illustrations are included to aid in depicting the points being made. In addition, the material draws upon the background and research literature indicated in Section 1 and combines with the material outlined in Section 2 on the Levels of Autonomy of AI Legal Reasoning.

### 3.1 AAI Overview

There is a synergistic aspect to antitrust and the field of Artificial Intelligence. In some respects, antitrust can be applied to AI, while in other respects AI can be applied to antitrust.

See **Figure B-1**.

As shown, there is a cyclical way to cast the roles of antitrust and AI. You can assert that in certain ways there is the impact of antitrust upon AI, or more properly phrased the advent of AI. You can also assert that in particular ways there is an impact of AI upon antitrust, primarily regarding the use or application of AI in antitrust efforts.

Let’s consider some examples associated with these in-common interactions.

See **Figure B-2**.

As an example of antitrust being applied to AI, consider that as discussed in Section 1 there is ample debate about the notion that businesses will make use of AI and thus potentially spur antitrust behaviors. Meanwhile, the counterargument is that AI used by businesses will disrupt the chances of undertaking antitrust behaviors. Of course, the reality is likely going to be that AI has a dual impact, for which in some instances antitrust behaviors will be sparked while in other instances antitrust behaviors will be lessened or ostensibly subverted via the use of AI in a marketplace.

For further illustration, see **Figure B-3**.

Another example of antitrust impacting or being applied to AI consists of the qualm that perhaps antitrust enforcement will dampen AI progress. This was also discussed in Section 1. Again, the impact of antitrust on AI does not necessarily need to be in one direction only. There is a possibility that antitrust enforcement, or the lack thereof, could accelerate the progress of AI. Likely, the use of antitrust enforcement is bound to have both an encouraging effect on the AI innovation ecosystem and simultaneously a dampening effect, dependent upon how the antitrust efforts are guided and utilized.

For further illustration, see **Figure B-4**.

In terms of applying AI to antitrust, this is the mainstay of the rest of this discussion and consists of establishing that AI can and will undoubtedly be integrated into antitrust efforts.

See **Figure B-5**.

This is coined as the use of AI to increase or bolster antitrust vigilance.

As an aside on terminology, typically, the process of antitrust efforts are usually referred to as antitrust *enforcement*. Here, we use instead the phrasing of antitrust *vigilance*.

The reason or rationale that antitrust vigilance is used rather than "enforcement" involves several significant points.

First, the enforcement of antitrust is oftentimes perceived as the latter part of the overall lifecycle associated with antitrust efforts (the lifecycle will be identified and explored in a moment). The application of AI to antitrust efforts will not be confined to only the enforcement portion of the lifecycle and will instead be applicable throughout the entire lifecycle. Thus, a sound basis for avoiding referring to the application of AI as to antitrust enforcement is to avoid the misconception that AI would only be applied to a segment or subset of the lifecycle.

Secondly, there is a connotation of the word "enforcement" that seems ominous and conjures imagery of an onerous nature. In theory, enforcement is merely the act of enforcing or ensuring that the antitrust laws are being legally obeyed. Nonetheless, there is a stigma associated with the notion of enforcement. If the application of AI to antitrust efforts is to take hold, there is a chance that a backlash could develop simply on the basis that it would seem unfair or unreasonable to use AI as a form of an *enforcer* per se. Though indeed AI is going to be used to aid in enforcing the antitrust laws, there would be a too easy assumption that AI is somehow overstepping appropriate bounds and being used in some revengeful or overzealous way. We avoid this confusion by referring to vigilance instead.

That being said, do not overstep this indication by also assuming that the use of AI might not be in fact overzealously utilized and in some respects become a kind of doomsday antitrust enforcer. There is a quite real possibility that AI if poorly or inadequately devised, could become an antitrust zealot that overreaches. To that degree, the forewarning of the word "enforcer" and its connotations does provide some benefit for use, though on the whole, it seems prudent for now to refer to these matters as one of vigilance.

In short, the phrasing of antitrust vigilance does away with those aforementioned qualms about the use of the word enforcement. AI is going to increase antitrust vigilance, for which the "enforcement" element is encompassed.

### 3.2 Antitrust Winnowing Funnel

Another facet of noteworthiness about antitrust and the vigilance process consists of the antitrust winnowing funnel.

See **Figure B-6**.

As shown, there is a funnel that starts in a wide manner and gradually winnows antitrust candidates until there is a narrower set considered viable for antitrust enforcement.

There is a continual effort of scrutinizing the marketplace for potential antitrust behaviors. This search is ongoing and wide. The odds are that most of the suspected or considered antitrust targets will fall out of the funnel due to a lack of sufficiently credible indication of their alleged antitrust deeds.

At each step forward in the funnel, those candidate firms that are seemingly more likely to have performed antitrust behavior will presumably be increasingly examined and assessed. In theory, only those firms meeting some level of threshold or degree will continue into the funnel to the point of actual enforcement.

This brings up via implication that there is a potential lifecycle associated with these antitrust pursuits, which is indeed the case and will be explored in the next subsection.

### 3.3 AAI Vigilance Lifecycle

There is an antitrust vigilance lifecycle, and it provides crucial insight about the nature of the antitrust processes and also serves as a means to overlay the application of AI onto the antitrust realm.

See **Figure B-7**.

As shown, there are six key phases (some refer to these as stages, which, in our view, can be interchangeably stated):

1. Detect
2. Assess
3. Investigate
4. Recommend
5. Prosecute
6. Implement

The Detect phase entails the widest part of the antitrust winnowing funnel and consists of an ongoing and persistent antitrust pursuit by observing marketplaces for signs of antitrust behaviors. There is plenty of voluminous signals and indications that need to be detected and distilled. This also has to be taking place persistently since the marketplace itself is dynamic and ever-changing.

For those spotted or anticipated potential antitrust behaviors, the next phase comes into play, namely the Assess phase. An assessment requires additional resources and therefore should be undertaken only when the Detect has identified viable candidates or targets for closer inspection.

Out of the assessment phase, there will be antitrust targets or candidates that are ascertained as worthwhile for even greater scrutiny and will therefore be funneled into the Investigate phase.

For underlying details associated with these processes and phases, the United States Department of Justice (DOJ) *Antitrust Division Manual* [82] provides an essential grounding in the details of these antitrust pursuit activities.

Based on the outcome of the Investigate phase, there will be some candidates or targets that are funneled into the Recommend phase. This will then consist of a formal recommendation for a civil or criminal case to be opened and undertaken.

Out of the recommended set, there will be some candidates or targets that are then prosecuted, thus the Prosecute phase. Finally, depending upon the outcome of the prosecution, there is likely to be a need to implement the result, doing so via monitoring for compliance and also taking added follow-up action if compliance falters or is not observed.

These six phases can be overlaid onto the antitrust winnowing funnel.

See **Figure B-8**.

We can now more broadly discuss the nature of AI being applied to antitrust vigilance, having set the table, as it were, by having established a viable context for understanding the fuller picture of the matter at hand.

Various aspects of AI will be applied to each of the six phases of the antitrust vigilance lifecycle.

Not all of the lifecycle phases will be injected with AI in the same manner and nor at the same pace. This is an important point. In particular, some researchers have blandly indicated that AI will be applied to antitrust efforts, though that is a quite overreaching indication and treats the antitrust process as a monolith. Similarly, in an inappropriate manner, AI is treated as a monolith, as though the mere reference to AI is tantamount to having some particular meaning, when in fact the notion of AI is an umbrella or collective of various technologies.

See **Figure B-9**.

As shown, AI can be considered as consisting of these overarching areas of technologies:

- Machine Learning (ML)
- Knowledge-Based Systems (KBS)
- Natural Language Processing (NLP)
- Computer Vision (CV)
- Robotics/Autonomy
- Common-Sense Reasoning
- Other Technologies

Each of those distinct areas of AI technologies will be gradually applied to antitrust, doing so across the antitrust vigilance lifecycle.

A grid is envisioned that would encompass the AI Technologies aligned with each of the six phases, such that it would be feasible to indicate the depth of coverage by each of the AI Technologies for each of the respective six phases. Such a grid would be updated over time as the infusion of AI into the antitrust vigilance lifecycle matures.

Speaking of the maturity aspects, the inclusion of AI into the antitrust vigilance lifecycle will significantly different depending upon the level of autonomy associated with AI.

In Section 2, a framework was provided to depict the levels of autonomy associated with AI-based legal reasoning. This provides a basis for next exploring the application of AI to antitrust in the context of the maturation of AI across the levels of autonomy.

### 3.4 AAI and AILR

The nature and capabilities of applied AAI will vary across the Levels of Autonomy for AI Legal Reasoning.

Refer to **Figure B-10**.

As indicated, applied AAI becomes increasingly more sophisticated and advanced as the AI Legal Reasoning increases in capability. To aid in typifying the differences between each of the Levels in terms of the incremental advancement of applied AAI, the following phrasing is used:

- Level 0: **n/a**
- Level 1: **Rudimentary (simplistic)**
- Level 2: **Complex (simplistic)**
- Level 3: **Symbolic Intermixed**
- Level 4: **Domain Incisive**
- Level 5: **Holistic Incisive**
- Level 6: **Pansophic Incisive**

Briefly, each of the levels is described next.

At Level 0, there is an indication of “n/a” at Level 0 since there is no AI capability at Level 0 (the *No Automation* level of the LoA).

At Level 1, the LoA is *Rudimentary (simplistic)* and this can be used to undertake applied AAI though it is rated or categorized as being rudimentary and making use of relatively simplistic calculative models and formulas. Thus, this is coined as “Rudimentary (Simplistic).”

At Level 2, the LoA is *Advanced Assistance Automation* and the applied AAI is coined as “Complex (Simplistic),” which is indicative of AAI being performed in a more advanced manner than at Level 1. This consists of complex statistical methods such as those techniques mentioned in Section 1 of this paper. To date, most of the research and practical use of applied AAI has been within Level 2. Future efforts are aiming at Level 3 and above.

At Level 3, the LoA is *Semi-Autonomous Automation* and the applied AAI is coined as “Symbolic Intermixed,” which can undertake AAI at an even more advanced capacity than at Level 2. Recall, in Level 2, the focus tended to be on traditional numerical formulations, albeit sophisticated in the use of statistical models. In Level 3, the symbolic capability is added and fostered, including at times acting in a hybrid mode with the conventional numerical and statistical models. Generally, the work at Level 3 to-date has primarily been experimental, making use of exploratory prototypes or pilot efforts.

At Level 4, the LoA is *AILR Domain Autonomous* and the applied AAI coined as "Domain Incisive," meaning that this can be used to perform AAI within particular specialties of domains or subdomains of the legal field but does not necessarily cut across the various domains and is not intended to be able to do so. The capacity is done in a highly advanced manner, incorporating the Level 3 capabilities, along with exceeding those levels and providing a more fluent and capable perceptive means.

At Level 5, the LoA is *AILR Fully Autonomous,* and the applied coined as "Holistic Incisive," meaning that the use of AAI can go across all domains and subdomains of the legal field. The capacity is done in a highly advanced manner, incorporating the Level 4 capabilities, along with exceeding those levels and providing a more fluent and capable perceptive means.

At Level 6, the LoA is *AILR Superhuman Autonomous*, which as a reminder from Section 2 is not a capability that exists and might not exist, though it is included as a provision in case such a capability is ever achieved. In any case, the applied AAI at this level is considered "Pansophic Incisive" and would encapsulate the Level 5 capabilities, and then go beyond that in a manner that would leverage the AI superhuman capacity.

## 4 Additional Considerations and Future Research

As earlier indicated, efforts to undertake the antitrust lifecycle have historically been performed by human hand and cognition, and only thinly aided in more recent times by the use of computer-based approaches.

Advances in Artificial Intelligence (AI) involving especially Natural Language Processing (NLP) and Machine Learning (ML) are increasingly bolstering how automation can systematically aid antitrust efforts. This research paper has examined the evolving infusion of AI into AAI, along with showcasing how the Levels of Autonomy (LoA) of AI Legal Reasoning (AILR) will impact this application.

Artificial Intelligence (AI) based approaches have been increasingly utilized and will undoubtedly have a pronounced impact on how antitrust is performed and its use in the practice of law, which will inevitably also have an impact upon theories of the law.

Future research is needed to explore in greater detail the manner and means by which AI-enablement will occur in the law along with the potential for both positive and adverse consequences. Autonomous AILR is likely to materially impact the effort, theory, and practice of AAI, including as a minimum playing a notable or possibly even pivotal role in such advancements.

## About the Author

Dr. Lance Eliot is the Chief AI Scientist at Techbrium Inc. and a Stanford Fellow at Stanford University in the CodeX: Center for Legal Informatics. He previously was a professor at the University of Southern California (USC) where he headed a multi-disciplinary and pioneering AI research lab. Dr. Eliot is globally recognized for his expertise in AI and is the author of highly ranked AI books and columns.

**Figure A-1**

## AI & Law: Levels of Autonomy For AI Legal Reasoning (AILR)

| Level | Descriptor | Examples | Automation | Status |
|---|---|---|---|---|
| 0 | No Automation | Manual, paper-based (no automation) | None | De Facto - In Use |
| 1 | Simple Assistance Automation | Word Processing, XLS, online legal docs, etc. | Legal Assist | Widely In Use |
| 2 | Advanced Assistance Automation | Query-style NLP, ML for case prediction, etc. | Legal Assist | Some In Use |
| 3 | Semi-Autonomous Automation | KBS & ML/DL for legal reasoning & analysis, etc. | Legal Assist | Primarily Prototypes & Research Based |
| 4 | AILR Domain Autonomous | Versed only in a specific legal domain | Legal Advisor (law fluent) | None As Yet |
| 5 | AILR Fully Autonomous | Versatile within and across all legal domains | Legal Advisor (law fluent) | None As Yet |
| 6 | AILR Superhuman Autonomous | Exceeds human-based legal reasoning | Supra Legal Advisor | Indeterminate |

*Figure 1: AI & Law - Autonomous Levels by Rows* *Source Author: Dr. Lance B. Eliot*

V1.3

**Figure A-2**

**AI & Law: Levels of Autonomy For AI Legal Reasoning (AILR)**

| | Level 0 | Level 1 | Level 2 | Level 3 | Level 4 | Level 5 | Level 6 |
|---|---|---|---|---|---|---|---|
| **Descriptor** | No Automation | Simple Assistance Automation | Advanced Assistance Automation | Semi-Autonomous Automation | AILR Domain Autonomous | AILR Fully Autonomous | AILR Superhuman Autonomous |
| **Examples** | Manual, paper-based (no automation) | Word Processing, XLS, online legal docs, etc. | Query-style NLP, ML for case prediction, etc. | KBS & ML/DL for legal reasoning & analysis, etc. | Versed only in a specific legal domain | Versatile within and across all legal domains | Exceeds human-based legal reasoning |
| **Automation** | None | Legal Assist | Legal Assist | Legal Assist | Legal Advisor (law fluent) | Legal Advisor (law fluent) | Supra Legal Advisor |
| **Status** | De Facto – In Use | Widely In Use | Some In Use | Primarily Prototypes & Research-based | None As Yet | None As Yet | Indeterminate |

*Figure 2: AI & Law - Autonomous Levels by Columns* *Source Author: Dr. Lance B. Eliot*

V1.3

**Figure B-1**

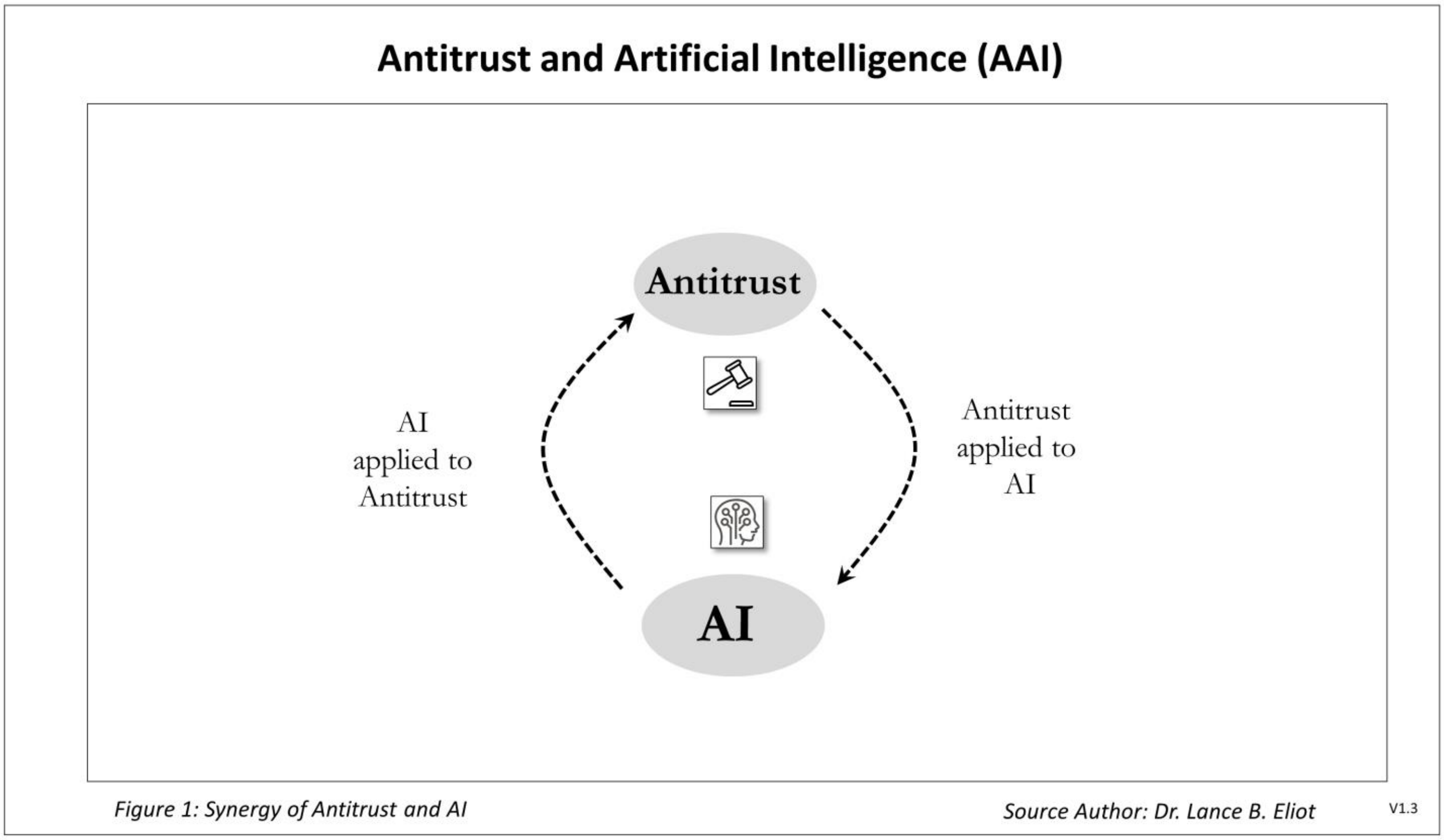

*Figure 1: Synergy of Antitrust and AI*

**Figure B-2**

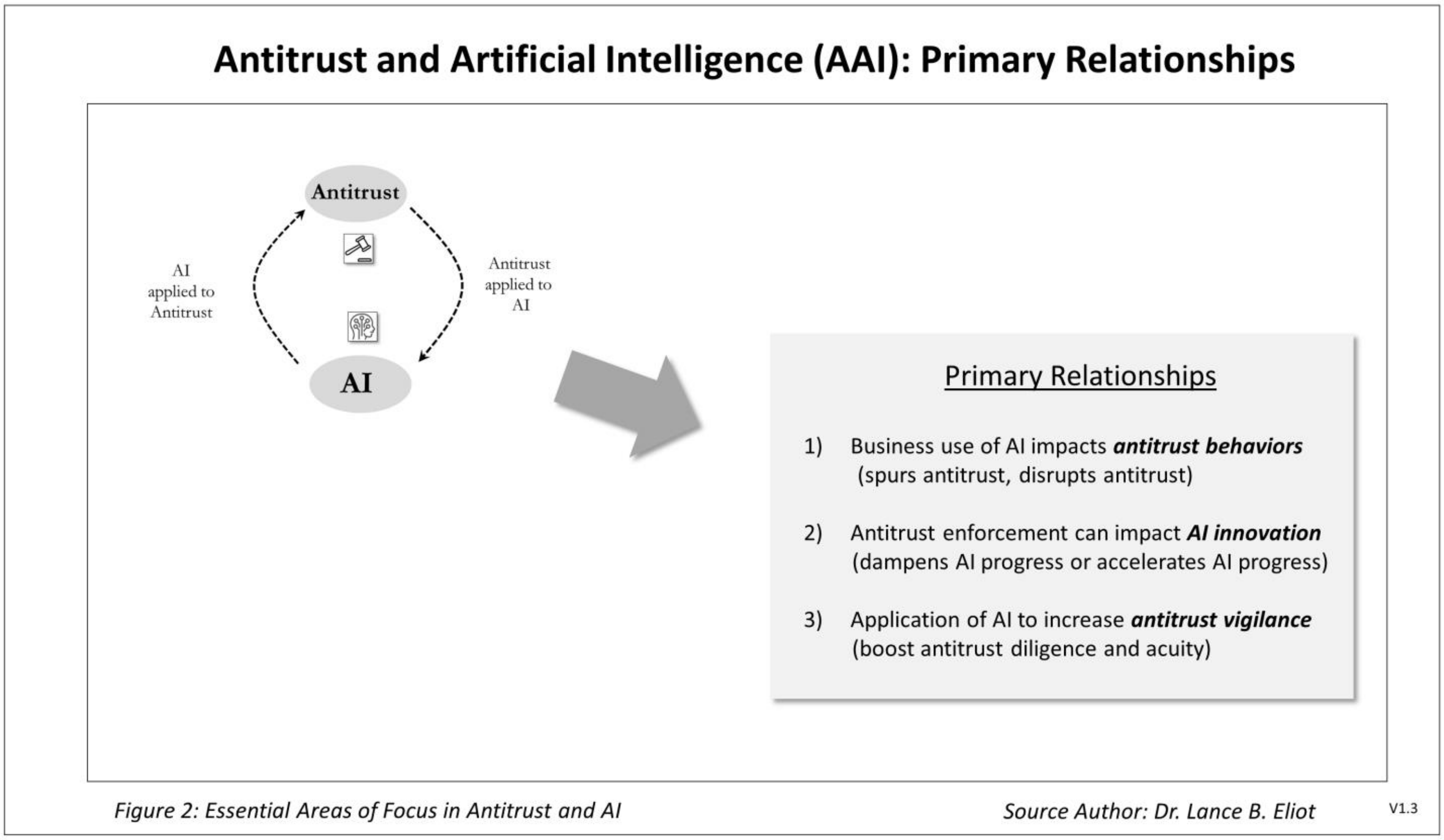
Antitrust and Artificial Intelligence (AAI): Primary Relationships
Antitrust
AI
AI applied to Antitrust
Antitrust applied to AI
Primary Relationships
1) Business use of AI impacts antitrust behaviors (spurs antitrust, disrupts antitrust)
2) Antitrust enforcement can impact AI innovation (dampens AI progress or accelerates AI progress)
3) Application of AI to increase antitrust vigilance (boost antitrust diligence and acuity)
Figure 2: Essential Areas of Focus in Antitrust and AI
Source Author: Dr. Lance B. Eliot
V1.3

**Figure B-3**

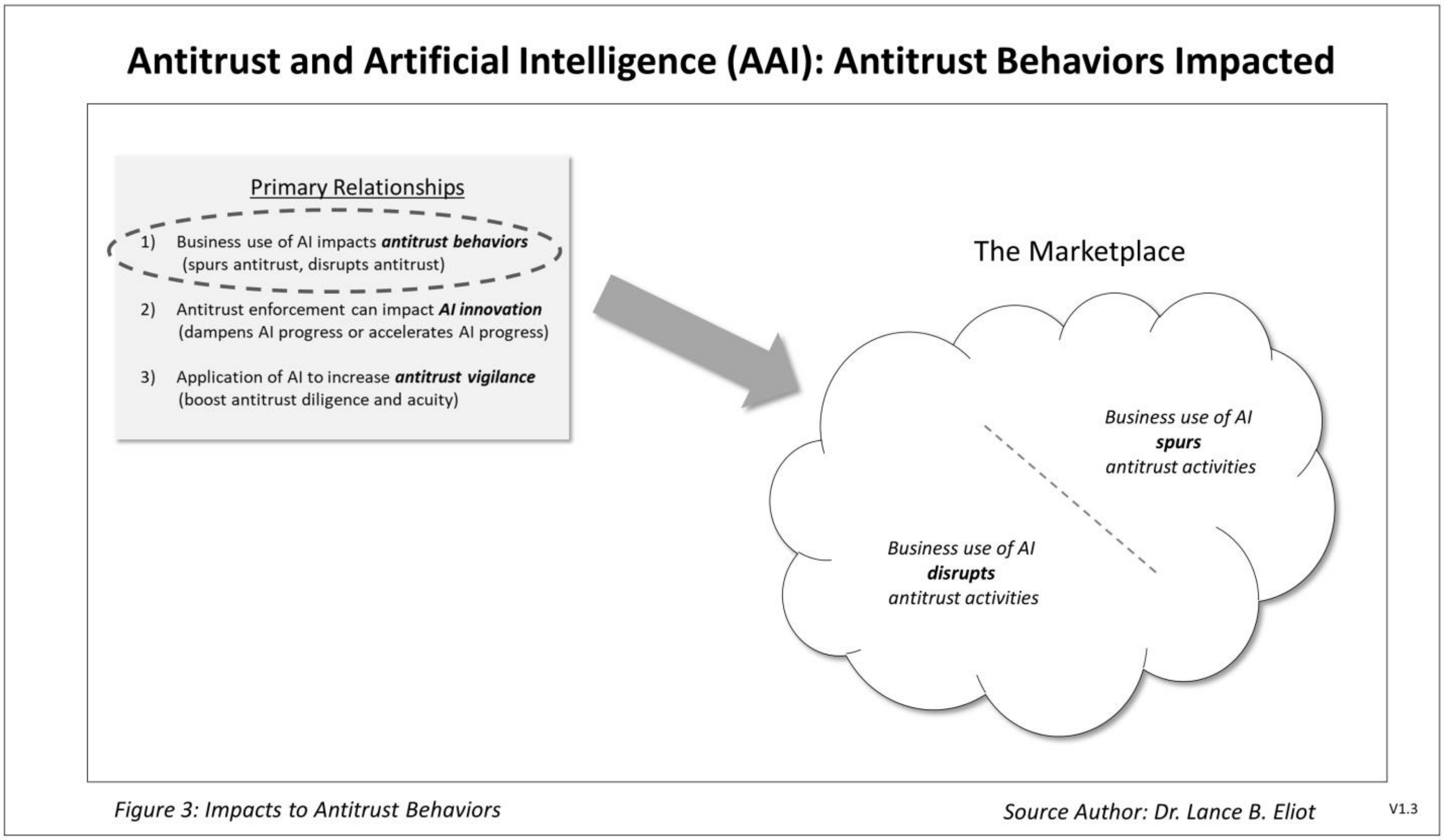

*Figure 3: Impacts to Antitrust Behaviors*

**Figure B-4**

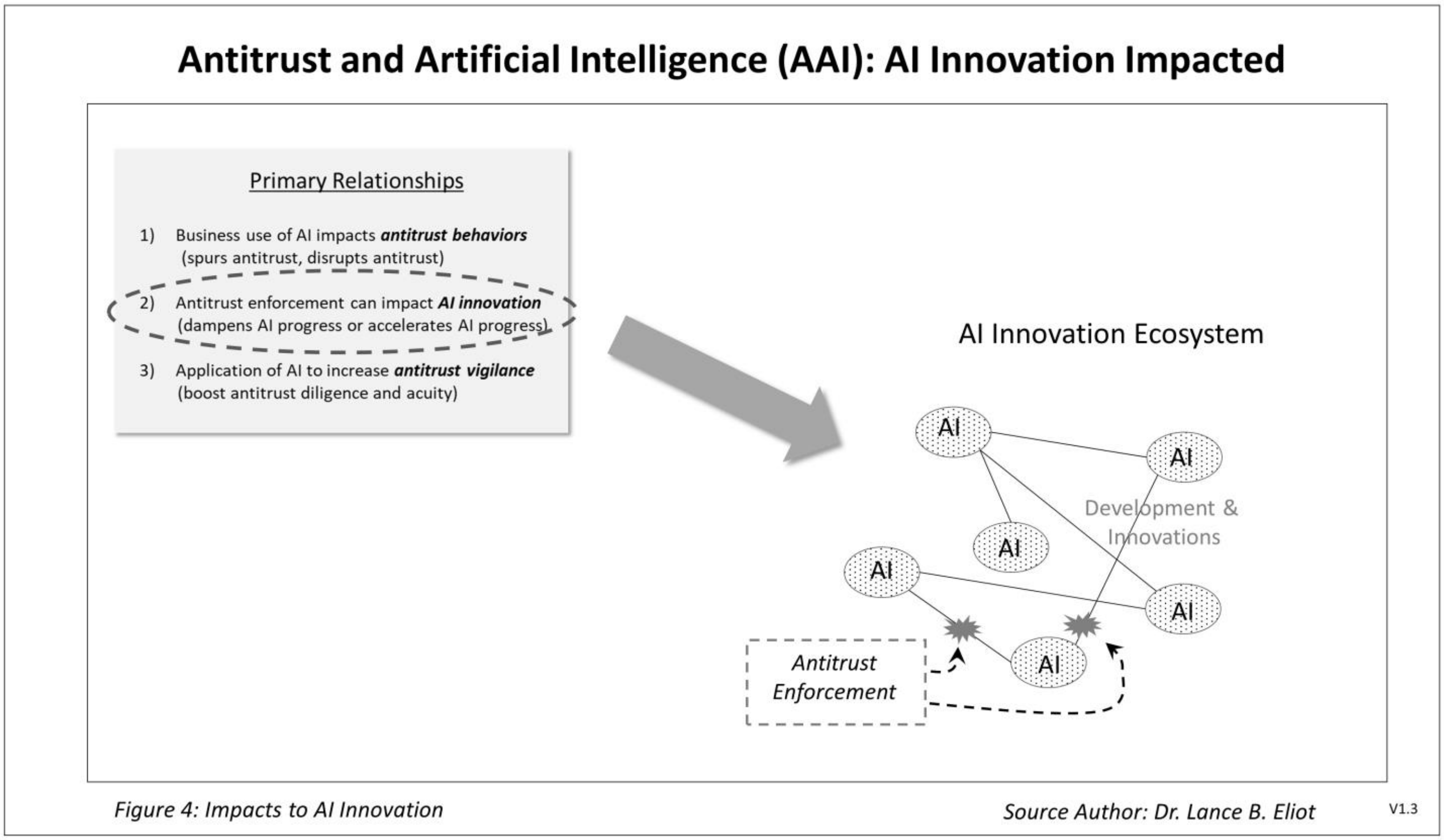
Antitrust and Artificial Intelligence (AAI): AI Innovation Impacted
Primary Relationships
1) Business use of AI impacts antitrust behaviors (spurs antitrust, disrupts antitrust)
2) Antitrust enforcement can impact AI innovation (dampens AI progress or accelerates AI progress)
3) Application of AI to increase antitrust vigilance (boost antitrust diligence and acuity)
AI Innovation Ecosystem
AI
AI
AI
AI
AI
AI
Development & Innovations
Antitrust Enforcement
Figure 4: Impacts to AI Innovation
Source Author: Dr. Lance B. Eliot
V1.3

**Figure B-5**

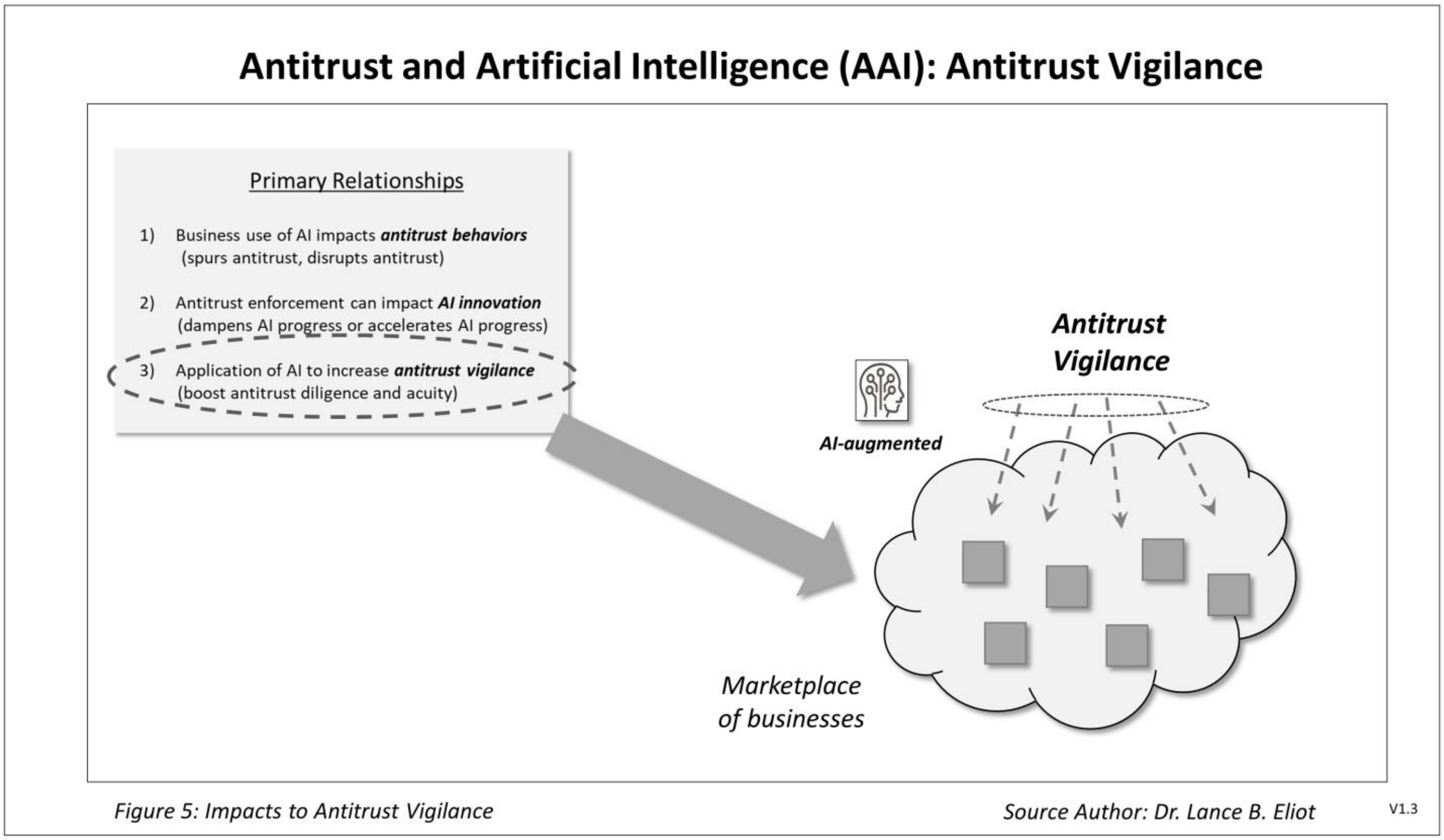
Antitrust and Artificial Intelligence (AAI): Antitrust Vigilance
Primary Relationships
1) Business use of AI impacts antitrust behaviors (spurs antitrust, disrupts antitrust)
2) Antitrust enforcement can impact AI innovation (dampens AI progress or accelerates AI progress)
3) Application of AI to increase antitrust vigilance (boost antitrust diligence and acuity)
Antitrust Vigilance
AI-augmented
Marketplace of businesses
Figure 5: Impacts to Antitrust Vigilance
Source Author: Dr. Lance B. Eliot
V1.3

**Figure B-6**

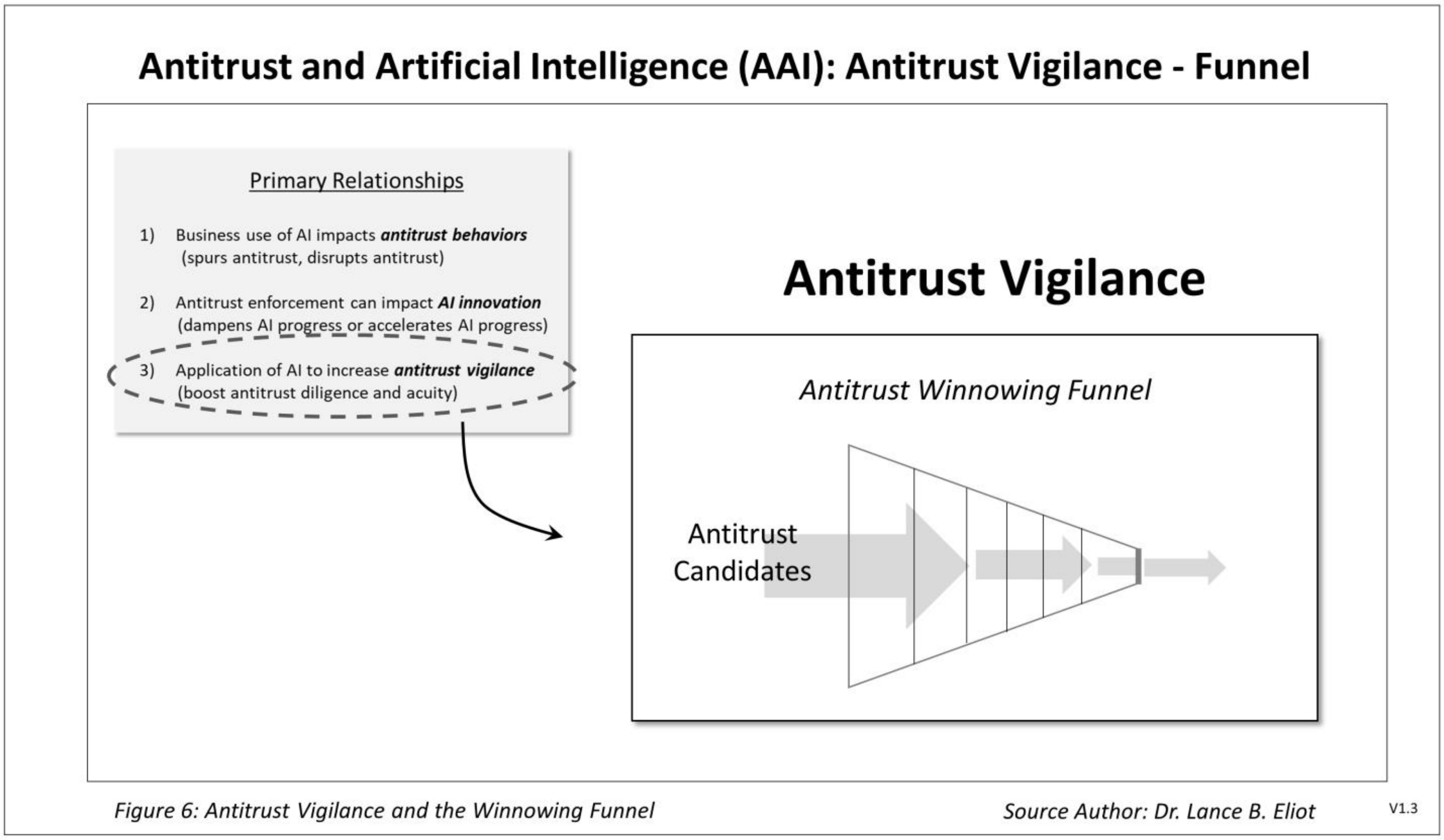

*Figure 6: Antitrust Vigilance and the Winnowing Funnel*

**Figure B-7**

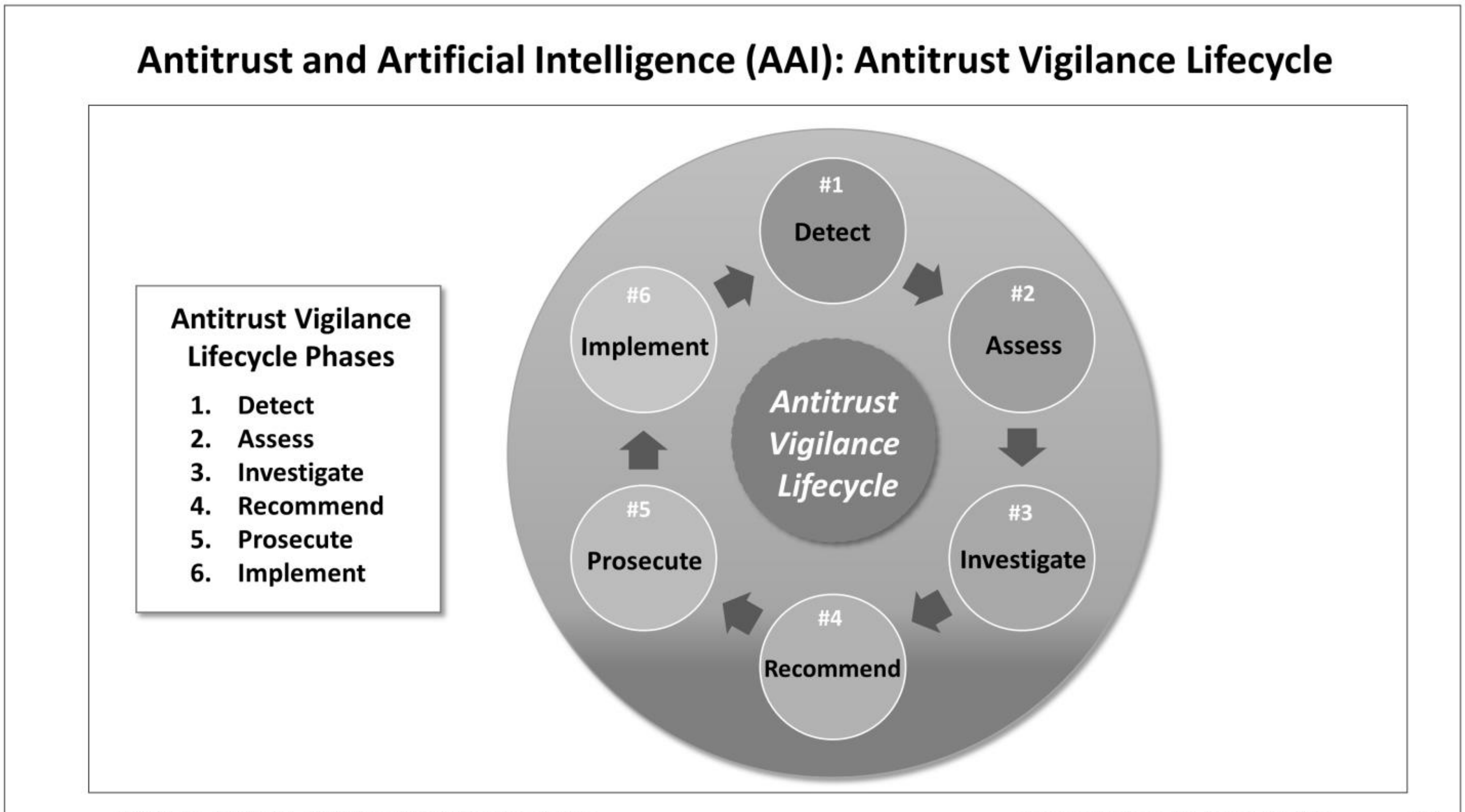

*Figure 7: Antitrust Vigilance Lifecycle and Phases* *Source Author: Dr. Lance B. Eliot*

**Figure B-8**

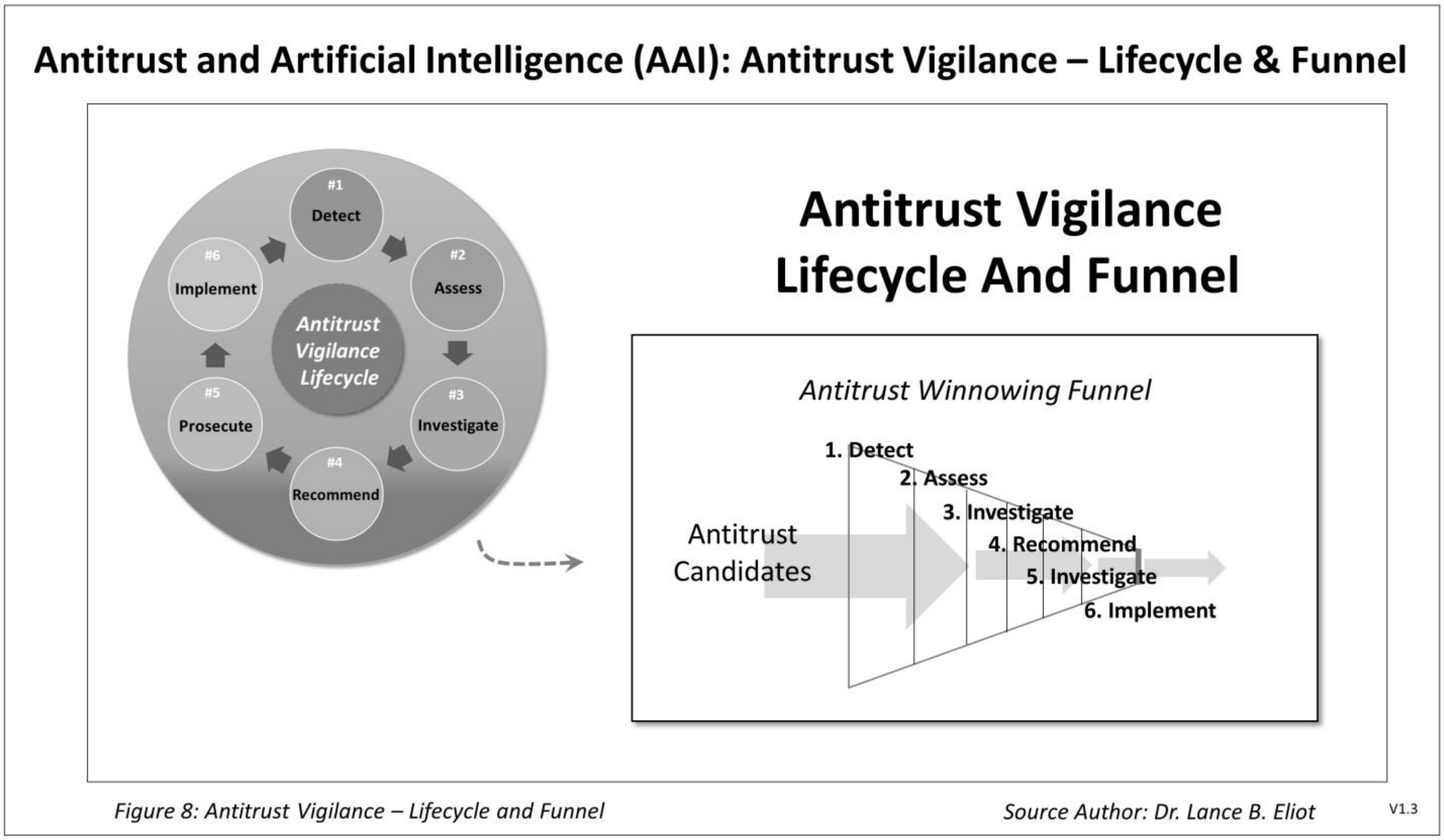

*Figure 8: Antitrust Vigilance – Lifecycle and Funnel*

**Figure B-9**

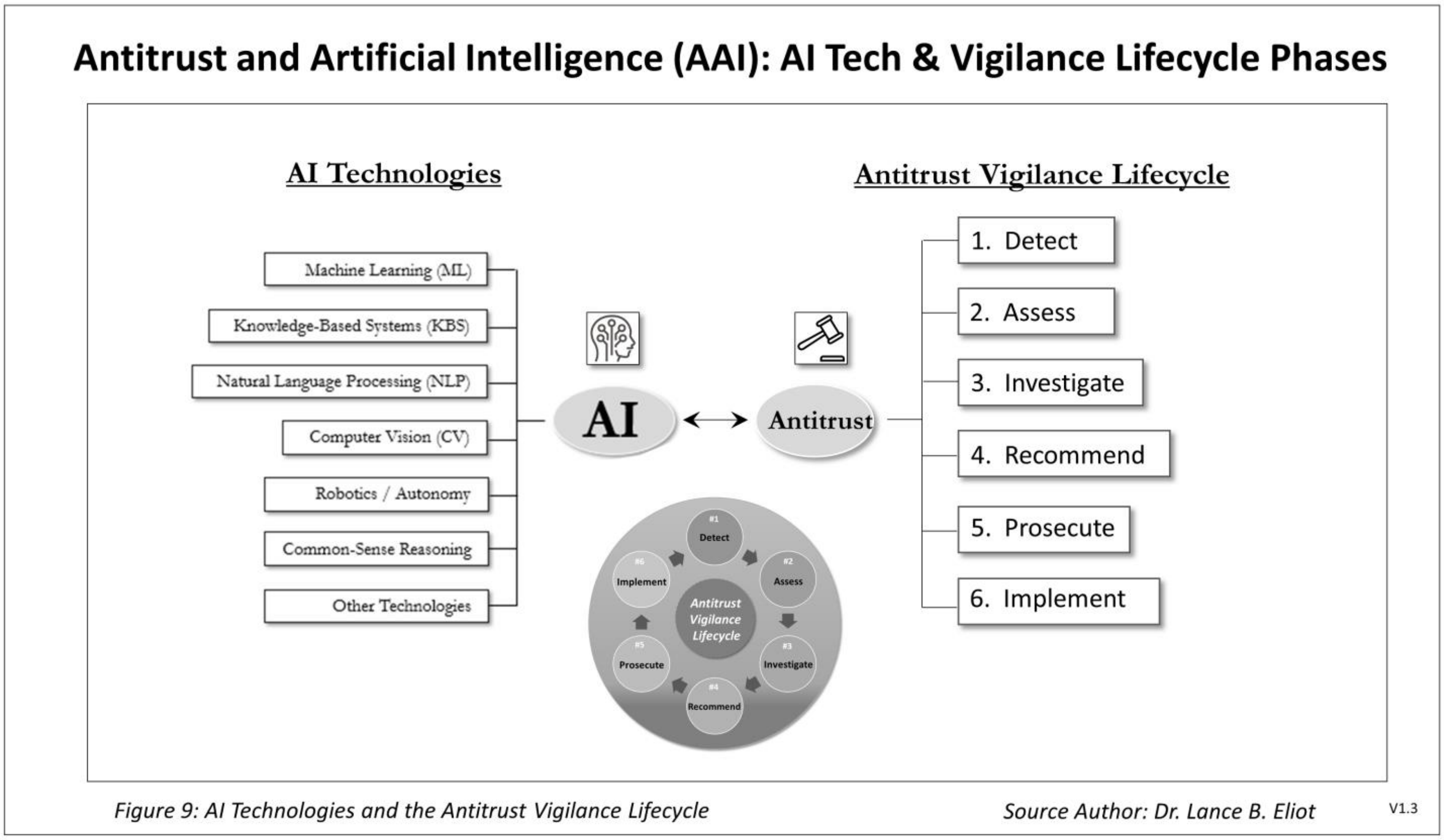

*Figure 9: AI Technologies and the Antitrust Vigilance Lifecycle*

**Figure B-10**

## Antitrust and AI: Levels of Autonomy Of AI Legal Reasoning (AILR)

| | Level 0 | Level 1 | Level 2 | Level 3 | Level 4 | Level 5 | Level 6 |
|---|---|---|---|---|---|---|---|
| **Descriptor** | No Automation | Simple Assistance Automation | Advanced Assistance Automation | Semi-Autonomous Automation | AILR Domain Autonomous | AILR Fully Autonomous | AILR Superhuman Autonomous |
| **Examples** | Manual, paper-based (no automation) | Word Processing, XLS, online legal docs, etc. | Query-style NLP, ML for case prediction, etc. | KBS & ML/DL for legal reasoning & analysis, etc. | Versed only in a specific legal domain | Versatile within and across all legal domains | Exceeds human-based legal reasoning |
| **Automation** | None | Legal Assist | Legal Assist | Legal Assist | Legal Advisor (law fluent) | Legal Advisor (law fluent) | Supra Legal Advisor |
| **Status** | De Facto – In Use | Widely In Use | Some In Use | Primarily Prototypes & Research-based | None As Yet | None As Yet | Indeterminate |
| **AI for Antitrust Vigilance** | *n/a* | *Rudimentary (simplistic)* | *Complex (simplistic)* | *Symbolic Intermixed* | *Domain Incisive* | *Holistic Incisive* | *Pansophic Incisive* |

*Figure 10: Antitrust and AI - Autonomous Levels of AILR by Columns* *Source Author: Dr. Lance B. Eliot*

V1.3